\documentclass[12pt,a4paper]{article}
\usepackage{amssymb,amsmath}
\usepackage[dvips]{lscape,graphicx}

\newcommand{\bc}{\begin{center}}
\newcommand{\ec}{\end{center}}
\newcommand{\bd}{\begin{displaymath}}
\newcommand{\ed}{\end{displaymath}}
\newcommand{\be}{\begin{equation}}
\newcommand{\ee}{\end{equation}}
\newcommand{\ba}{\begin{array}}
\newcommand{\ea}{\end{array}}
\newcommand{\bt}{\begin{tabular}}
\newcommand{\et}{\end{tabular}}

\newcommand{\ds}{\displaystyle}

\begin{document}

\begin{titlepage}

\vspace*{-15mm}
\begin{flushright}
{CERN-PH-TH/2007-189}\\
\end{flushright}
\vspace*{5mm}

\begin{center}
{
\LARGE
Fixed point scenario in the Two Higgs Doublet Model inspired by degenerate vacua.}\\[8mm]
{\large
C.~D.~Froggatt$^a$, R.~Nevzorov$^a$\,${}^{1}$,
H.~B.~Nielsen$^b$, D.~Thompson$^a$\\[3mm]
\itshape{$^a$ Department of Physics and Astronomy,}\\[2mm]
\itshape{Glasgow University, Glasgow, Scotland}\\[2mm]
\itshape{$^b$ The Niels Bohr Institute, Copenhagen, Denmark}
}\\[1mm]
\end{center}
\vspace*{0.75cm}

\begin{abstract}{
\noindent 
We consider the renormalisation group flow of Higgs and
Yukawa couplings within the simplest non--supersymmetric two Higgs
doublet extension of the Standard Model (SM). In this model the
couplings are adjusted so that the multiple point principle (MPP)
assumption, which implies the existence of a large set of
degenerate vacua at some high energy scale $\Lambda$, is realised.
When the top quark Yukawa coupling at the scale $\Lambda$ is
large, the solutions of RG equations in this MPP inspired 2 Higgs
Doublet Model (2HDM) converge to quasi--fixed points. We analyse
the Higgs spectrum and couplings in the quasi--fixed point
scenario and compute a theoretical upper bound on the lightest
Higgs boson mass. When the scale $\Lambda$ is low, the coupling of
the SM--like Higgs scalar to the top quark can be significantly
larger in the considered model than in the SM, resulting in the
enhanced production of Higgs bosons at the LHC.}
\end{abstract}

\vspace{6cm} 
\footnoterule{\noindent${}^{1}$
On leave of absence from the Theory Department, ITEP, Moscow, Russia}

\end{titlepage}

\newpage
\section{Introduction}

A quasi--fixed point solution
\cite{Hill:1980sq}--\cite{Hill:1985tg} is one of the most
spectacular features of the renormalisation group (RG) equations.
The existence of a quasi--fixed point implies that the solutions
of the RG equations, corresponding to a range of different initial
values of fundamental parameters at a high energy scale, are
focused in a narrow interval in the infrared region.
This allows us to get some predictions for couplings and physical
observables at low energy scales. However such predictions are not
always compatible with the existing experimental data. For
example, within the Standard Model (SM) the quasi--fixed point
solution leads to an unacceptably large mass for the top--quark
which disagrees with the results of experimental measurements
obtained at FNAL.

This problem can be overcome within supersymmetric (SUSY) and
non--supersymmetric two Higgs doublet extensions of the SM. The
most general renormalizable scalar potential of the model
involving two Higgs doublets is given by \be
\begin{array}{c}
V_{eff}(H_1, H_2) = m_1^2 H_1^{\dagger}H_1 + m_2^2 H_2^{\dagger}H_2 -
\biggl[m_3^2 H_1^{\dagger}H_2+h.c.\biggr]+\\[3mm]
\ds\frac{\lambda_1}{2}(H_1^{\dagger}H_1)^2 +
\frac{\lambda_2}{2}(H_2^{\dagger}H_2)^2
\lambda_3(H_1^{\dagger}H_1)(H_2^{\dagger}H_2) + \lambda_4|H_1^{\dagger}H_2|^2+\\[3mm]
\ds + \biggl[\frac{\lambda_5}{2}(H_1^{\dagger}H_2)^2 + \lambda_6 (H_1^{\dagger}H_1)(H_1^{\dagger}H_2)+
\lambda_7 (H_2^{\dagger}H_2)(H_1^{\dagger}H_2)+h.c. \biggr]
\end{array}
\label{2hdm1} \ee where
$H_n=\left(\chi^+_n,\,\ds\frac{1}{\sqrt{2}}(H_n^0+iA_n^0)\right)$,
$n=1,2$. In the minimal supersymmetric standard model (MSSM) Higgs
self--couplings $\lambda_5$, $\lambda_6$ and $\lambda_7$ are zero
at the tree level while the values of $\lambda_1$, $\lambda_2$,
$\lambda_3$ and $\lambda_4$ are proportional to the gauge
couplings squared. After the inclusion of loop corrections all
possible Higgs self--couplings are generated and the values of the
$\lambda_i$ at the electroweak scale depend on the soft SUSY
breaking parameters.

In the non--supersymmetric two Higgs doublet extension of the SM
(2HDM) the Higgs self--couplings $\lambda_i$ and the mass terms
$m_i^2$ are arbitrary parameters. In order to suppress
non--diagonal flavour transitions in the 2HDM, a certain discrete
$Z_2$ symmetry is normally imposed. This $Z_2$ symmetry requires
the down-type quarks to couple to just one Higgs doublet, $H_1$
say, while the up-type quarks couple either to the same Higgs
doublet $H_1$ (Model I) or to the second Higgs doublet $H_2$
(Model II) but not both \cite{2hdm} \footnote{Due to the
invariance of the Lagrangian of the 2HDM under this symmetry the
leptons can only couple to one Higgs doublet as well, usually
chosen to be the same as the down-type quarks.}. The custodial
$Z_2$ symmetry forbids the mixing term $m_3^2 (H_1^{\dagger}H_2)$
and the Higgs self--couplings $\lambda_6$ and $\lambda_7$. But
usually a soft violation of the $Z_2$ symmetry by dimension--two
terms is allowed, since it does not induce Higgs--mediated
tree--level flavor changing neutral currents (FCNC).

At the physical minimum of the scalar potential (\ref{2hdm1}) the
neutral components of the Higgs doublets develop vacuum
expectation values $<H_1^0>=\ds\frac{v_1}{\sqrt{2}}$ and
$<H_2^0>=\ds\frac{v_2}{\sqrt{2}}$, breaking electroweak symmetry
and generating masses for the bosons and fermions. In the MSSM and
2HDM of type II, the induced running $t$--quark mass $m_t$ is
given by \be m_t(M_t)=\ds\frac{h_t(M_t) v}{\sqrt{2}}\sin\beta\, ,
\label{2hdm6} \ee where $M_t=171.4\pm 2.1$ GeV is the top quark
pole mass \cite{Brubaker:2006xn} and
$v=\sqrt{v_1^2+v_2^2}=246\,\mbox{GeV}$ is fixed by the Fermi
scale, while $\tan\beta=v_2/v_1$ remains arbitrary. Because
$\sin\beta$ can be considerably smaller than unity a
phenomenologically acceptable value of $m_t(M_t)$ can be obtained
even for $h_t(M_t)\gtrsim 1$, which is not the case in the SM
where such large values of the top quark Yukawa coupling have
already been ruled out. In the MSSM a broad class of solutions of
the RG equations converges to the quasi--fixed point which
corresponds to $\tan\beta\simeq 1.3-1.8$, resulting in a stringent
constraint on the lightest Higgs boson mass $m_h\lesssim 94\pm
5\,\mbox{GeV}$ \cite{mssm1}--\cite{mssm4}. Such a light Higgs
boson has already been excluded by LEP II data. But at large
$\tan\beta=50-60$ the solutions of the MSSM RG equations are
focused near another quasi--fixed point, which has not been ruled
out by LEP measurements. The RG flow of Yukawa couplings and the
particle spectrum in the vicinity of the MSSM quasi--fixed points
were discussed in \cite{mssm4}--\cite{mssm5}. The quasi--fixed
point scenario in the non--supersymmetric two Higgs doublet
extension of the SM was studied in \cite{Hill:1985tg},
\cite{Froggatt:1990wa}.

In this letter we consider the quasi--fixed point scenario within
a specific two Higgs doublet model obtained from the application
of the multiple point principle (MPP) to the 2HDM of type II. The
MPP postulates the existence of the maximal number of phases with
the same energy density allowed by a given theory \cite{mpp}.
Being applied to the 2HDM of type II, the multiple point principle
implies the existence of a large set of degenerate vacua at some
high energy scale $\Lambda$ (MPP scale). To ensure that the vacua
at the electroweak and MPP scales have the same vacuum energy
density, $\lambda_5$ must have zero value while
$\lambda_1(\Lambda)$, $\lambda_2(\Lambda)$, $\lambda_3(\Lambda)$
and $\lambda_4(\Lambda)$ obey two MPP conditions (see
\cite{mpp-2hdm}). Thus the MPP inspired 2HDM has less free
parameters than the 2HDM of type II and therefore can be
considered as a minimal non--supersymmetric two Higgs doublet
extension of the SM. Also it has recently been shown that the MPP
can be used to derive a softly broken custodial symmetry, which
suppresses FCNC and CP violating phenomena in the 2HDM
\cite{z2hdm}.

This letter is organised as follows. In the next section we
examine the RG flow of $h_t(\mu)$ and $\lambda_i(\mu)$ and
determine the position of the quasi--fixed points to which the
solutions of the RG equations approach when $h_t(\Lambda)\gtrsim 1$.
In section 3 the results obtained are used in an analysis of the
Higgs masses and couplings. We establish an upper bound on the
mass of the SM--like Higgs boson in the vicinity of the
quasi--fixed point and argue that the Higgs production cross
section at the LHC can be significantly larger in the considered
model as compared with the SM. 
Our results are summarised in section 4.

\section{RG flow of Higgs and Yukawa couplings}

Let us consider the running of Higgs and Yukawa couplings in the
framework of the MPP inspired 2HDM. At moderate values of
$\tan\beta$ ($\tan\beta\lesssim 10$), all Yukawa couplings except
the top quark one are negligibly small and can be safely ignored
in our analysis of the RG flow. As a consequence the RG equations
are simplified drastically and an exact analytic solution for
$h_t(\mu)$ may be obtained. It can be written as follows
\begin{equation}
\begin{gathered}
Y_t(\mu)=\frac{\dfrac{2 E(t)}{9 F(t)}}{1+\dfrac{2}{9 Y_t(\Lambda) F(t)}},\qquad
\tilde{\alpha}_i(\mu)=\frac{\tilde{\alpha}_i(\Lambda)}{1+b_i\tilde{\alpha}_i(\Lambda)\,t},\\
E(t)=\left[\frac{\tilde{\alpha}_3(\mu)}{\tilde{\alpha}_3(\Lambda)}\right]^{8/7}
\left[\frac{\tilde{\alpha}_2(\mu)}{\tilde{\alpha}_2(\Lambda)}\right]^{3/4}
\left[\frac{\tilde{\alpha}_1(\mu)}{\tilde{\alpha}_1(\Lambda)}\right]^{-17/84},
\quad F(t)=\int\limits_0^t E(\tau)d\tau,
\end{gathered}
\label{2hdm31}
\end{equation}
where the index $i$ varies from $1$ to $3$, $b_1=7$, $b_2=-3$,
$b_3=-7$, $t=\ln(\Lambda^2/\mu^2)$,
$\tilde{\alpha}_i(\mu)=\left(\dfrac{g_i(\mu)}{4\pi}\right)^2$ and
$Y_t(\mu)=\left(\dfrac{h_t(\mu)}{4\pi}\right)^2$. Here $g_i(\mu)$
are the gauge couplings of $U(1)_Y$, $SU(2)_W$ and $SU(3)_C$
interactions. If the MPP scale is relatively high and
$h^2_t(\Lambda)\gtrsim 1$ the second term in the denominator of
the expression describing the evolution of $Y_t(\mu)$ is much
smaller than unity at the electroweak scale. As a result the
dependence of $h_t^2(M_t)$ on its initial value $h_t^2(\Lambda)$
disappears and all solutions of the RG equation for the top quark
Yukawa coupling are concentrated in a narrow interval near the
quasi--fixed point \cite{Hill:1980sq}--\cite{Hill:1985tg}:
\begin{equation}
Y_\text{QFP}(M_t)=\dfrac{2\, E(t_0)}{9\,F(t_0)}\,,
\label{2hdm32}
\end{equation}
where $t_0=\ln(\Lambda^2/M_t^2)$. Formally a solution of this type
can be obtained in the limit when $Y_t(\Lambda)$ is infinitely
large. But in reality the convergence of the RG solutions to the
quasi--fixed point (\ref{2hdm32}) does not require extremely large
values of the top quark Yukawa coupling at the MPP scale if
$\Lambda$ is high enough. In Figs.~1a and 1b we plot the RG flow
of the top quark Yukawa coupling for different initial values at
the scale $\Lambda\simeq M_{Pl}$ and $\Lambda\simeq
10^{13}\,\mbox{GeV}$ respectively. One can see that in both cases
the solutions of the RG equations are focused in the close
vicinity of the quasi--fixed point at the electroweak scale if
$h_t^2(\Lambda)\gtrsim 1$ \footnote{The solutions of the RG
equations also converge to a quasi--fixed point at large
$\tan\beta=50-60$. However this quasi--fixed point scenario leads
to an unacceptably large $m_t(M_t)\gtrsim 200\,\mbox{GeV}$ (see
\cite{z2hdm}).}.

The convergence of the RG solutions to the quasi--fixed point
allows us to predict $h_t(M_t)$ for each fixed value of the MPP
scale. Then using Eq.~(\ref{2hdm6}) one can find the $\tan\beta$
that corresponds to the quasi--fixed point (\ref{2hdm32}). Here we
use the relationship between the $t$--quark pole ($M_t$) and
running ($m_t(\mu)$) masses \cite{mtMS}
\begin{equation}
m_t(M_t)=M_t\biggl[1-1.333\ds\,\frac{\alpha_s(M_t)}{\pi}-
9.125\left(\ds\frac{\alpha_s(M_t)}{\pi}\right)^2\biggr].
\label{2hdm35}
\end{equation}
to determine $m_t(M_t)$ within the $\overline{MS}$ scheme. We find
that in the two--loop approximation $m_t(M_t)\simeq 161.6\pm
2\,\mbox{GeV}$. In Table 1 we examine the dependence of the values
of $h_t(M_t)$ and $\tan\beta$ corresponding to the quasi--fixed
point (\ref{2hdm32}) on the MPP scale. From Table 1 it becomes
clear that $h_t(M_t)$ varies from $1.3$ to $2$ when the scale
$\Lambda$ changes from $M_{Pl}$ to $10\,\mbox{TeV}$. Because the
quasi--fixed point solution represents the upper bound on
$h_t(M_t)$, the value of $\tan\beta$ derived from
Eq.~(\ref{2hdm6}) should be associated with a lower bound on
$\tan\beta$. Then from Table 1 one can see that the lower limit on
$\tan\beta$ reduces from $1.1$ to $0.5$, when $\Lambda$ varies
from $M_{Pl}$ to $10\,\mbox{TeV}$.

It turns out that at large values of $h_t(\Lambda)\gtrsim 1.5$,
the allowed range of the Higgs self--couplings at the MPP scale is
quite narrow. Stringent constraints on $\lambda_i(\Lambda)$ come
from the MPP conditions. The MPP scale vacua
have small vacuum energy densities ($\ll \Lambda^4$), as needed to
achieve the degeneracy of these vacua and the physical one, only
if the Higgs self--couplings obey the MPP conditions \be
\lambda_3(\Lambda)=-\sqrt{\lambda_1(\Lambda)\lambda_2(\Lambda)}-\lambda_4(\Lambda)\,,
\label{2hdm7} \ee \be \ba{c}
\lambda_4^2(\Lambda)=\ds\frac{6h_t^4(\Lambda)\lambda_1(\Lambda)}
{(\sqrt{\lambda_1(\Lambda)}+\sqrt{\lambda_2(\Lambda)})^2}
-2\lambda_1(\Lambda)\lambda_2(\Lambda)\\[4mm]
\ds-\frac{3}{8}\biggl(3g_2^4(\Lambda)+2g_2^2(\Lambda)g_1^2(\Lambda)+g_1^4(\Lambda)\biggr)\,,
\ea \label{2hdm71} \ee where $\lambda_4(\Lambda)<0$. Thus, in
contrast to the 2HDM of type II, the Higgs self--couplings
$\lambda_3(\Lambda)$ and $\lambda_4(\Lambda)$ in the MPP inspired
two Higgs doublet extension of the SM are determined by
$\lambda_1(\Lambda)$, $\lambda_2(\Lambda)$ and $h_t(\Lambda)$.
These three parameters determine the RG flow of all the couplings
in the considered model. Since $\lambda_4(\Lambda)$ is a real
quantity, Eq.~(\ref{2hdm71}) limits the allowed range of
$\lambda_1(\Lambda)$ and $\lambda_2(\Lambda)$ from above. For
instance, when $\lambda_1(\Lambda)=\lambda_2(\Lambda)=\lambda_0$
the value of $\lambda_4^2(\Lambda)$ remains positive only if
$\lambda_0<\ds\frac{\sqrt{3}}{2}h_t^2(\Lambda)$.

The lower bound on the Higgs self--couplings originates from the
vacuum stability conditions: \be \lambda_1(\mu)>0,\qquad
\lambda_2(\mu)>0,\qquad
\tilde{\lambda}(\mu)=\sqrt{\lambda_1(\mu)\lambda_2(\mu)}+\lambda_3(\mu)+\min\{0,\lambda_4(\mu)\}>0.
\label{2hdm46} \ee The conditions (\ref{2hdm46}) must be fulfilled
everywhere from the electroweak scale to the MPP scale. Otherwise
another minimum of the Higgs effective potential with a huge and
negative vacuum energy density arises at some intermediate scale,
destabilising the physical and MPP scale vacua and preventing the
consistent realisation of the MPP in the 2HDM. The running of
gauge, Yukawa and Higgs couplings in the MPP inspired 2HDM is
described by a system of RG equations, which is basically the same
as in the 2HDM of type II but with $\lambda_5=0$. The set of
one--loop RG equations for the two Higgs doublet model with exact
and softly broken $Z_2$ symmetry can be found in
\cite{Hill:1985tg}, \cite{rg1}--\cite{rg}.

For the purposes of our RG studies it is convenient to define
\begin{equation}
\rho_i(\mu)=\dfrac{\lambda_i(\mu)}{g_3^2(\mu)}\,,\qquad
\rho_t(\mu)=\dfrac{h^2_t(\mu)}{g_3^2(\mu)}\,,\qquad
R_i(\mu)=\dfrac{\rho_i(\mu)}{\rho_t(\mu)}=\dfrac{\lambda_i(\mu)}{h_t^2(\mu)}.
\label{2hdm43}
\end{equation}
The vacuum stability constraints (\ref{2hdm46}) and the MPP
conditions (\ref{2hdm7})--(\ref{2hdm71}) confine the allowed range
of $R_i(\Lambda)$ in the vicinity of \be R_1=\dfrac{3}{4}\,,\qquad
R_2=\dfrac{\sqrt{65}-1}{8}\simeq 0.883\,,\qquad R_3=R_4=0\,,
\label{2hdm47} \ee which is a stable fixed point of the RG
equations in the gaugeless limit ($g_i=0$). Our numerical studies
show that for $\Lambda=M_{Pl}$ and $R_1(M_{Pl})=R_2(M_{Pl})=R_0$
the value of $R_0$ can vary only within a very narrow interval
from $0.79$ to $0.87$ if $h_t(\Lambda)\gtrsim 1.5$. Moreover the
allowed range of $R_0$ shrinks significantly when $h_t(\Lambda)$
increases. For $h_t(\Lambda)\gtrsim 2.5$ the value of $R_0$ can
only vary between $0.83$ and $0.87$. When the MPP scale decreases
the allowed range of $\lambda_1(\Lambda)$ and $\lambda_2(\Lambda)$
enlarges.

Because in the MPP inspired 2HDM the $R_i(\Lambda)$ are confined
near the fixed point (\ref{2hdm47}), the corresponding solutions
of the RG equations are attracted towards the invariant line that
joins the stable fixed point in the gaugeless limit to the
infrared stable fixed point
\begin{equation}
\rho_t=\dfrac{2}{9},\qquad \rho_1=0,\qquad \rho_2=\dfrac{\sqrt{689}-25}{36}\simeq 0.0347,
\qquad \rho_3=\rho_4=0\,,
\label{2hdm44}
\end{equation}
where all the solutions of the RG equations are concentrated when
the strong gauge coupling $g_3(\mu)$ approaches the Landau pole.
As a result at the electroweak scale the solutions of the RG
equations for the Higgs self--couplings are gathered in
the vicinity of the quasi--fixed point, which is an intersection
point of the invariant line and the Hill type effective surface
\cite{qfp}. Infrared fixed lines and surfaces as well as their
properties were studied in detail in \cite{infr-line}.

In Fig.2 we examine the RG running of the $\lambda_i(\mu)$. We set
$\Lambda$ equal to the Planck scale. Different curves in Fig.2
represent different solutions of the RG equations with boundary
conditions satisfying Eqs.~(\ref{2hdm7})--(\ref{2hdm71}) where we
keep $\lambda_1(\Lambda)=\lambda_2(\Lambda)=\lambda_0$. Because
there is a stringent correlation between $\lambda_0$ and
$h_t(\Lambda)$ we vary these couplings simultaneously, i.e. each
curve below the quasi--fixed point solution corresponds to a
particular set of $\lambda_0$ and $h_t(\Lambda)$ values. From
Fig.2 one can see that at low energies the solutions of the RG
equations for $\lambda_1(\mu)$, $\lambda_2(\mu)$ and
$\lambda_3(\mu)$ are focused in a narrow interval near their
quasi--fixed points. At the same time the solutions of the RG
equations for $\lambda_4(\mu)$ are attracted to the corresponding
quasi--fixed point rather weakly.

In Table 1 we specify the values of $\hat{\lambda}_i(M_t)$ to
which the solutions of the RG equations converge at large
$h_t(\Lambda)$. The set of $\hat{\lambda}_i(M_t)$ presented in
Table 1 is obtained for $h_t^2(\Lambda)=10$, $R_1(\Lambda)=0.75$
and $R_2(\Lambda)\simeq 0.883$. The other Higgs self--couplings
$\lambda_3(\Lambda)$ and $\lambda_4(\Lambda)$ are determined from
the MPP conditions (\ref{2hdm7})--(\ref{2hdm71}). In Table 1 we
present a few different sets of the Higgs self--couplings at the
electroweak scale that correspond to different choices of the
scale $\Lambda$ between $M_{Pl}$ and $10\,\mbox{TeV}$. The results
given in Table 1 demonstrate that the absolute values of
$\hat{\lambda}_i(M_t)$ increase as $\Lambda$ approaches the
electroweak scale. However the convergence of the Higgs
self--couplings to $\hat{\lambda}_i(M_t)$ becomes weaker as the
interval of evolution $t_0=\ln(\Lambda^2/M_t^2)$ shrinks. In
general the solutions of the RG equations for $\lambda_1(\mu)$ and
$\lambda_2(\mu)$ are attracted to their quasi--fixed points much
stronger than $\lambda_3(\mu)$ and $\lambda_4(\mu)$.

\section{Higgs masses and couplings}

Relying on the results of the analysis of the RG flow for the top
quark Yukawa and Higgs couplings one can explore the Higgs
spectrum in the MPP inspired 2HDM. The constraints on the Higgs
masses in the 2HDM with unbroken $Z_2$ symmetry have been examined
in a number of publications \cite{rg}, \cite{2hdm2}. The
theoretical restrictions on the mass of the SM--like Higgs boson
within the 2HDM with softly broken $Z_2$ symmetry were studied in
\cite{Kanemura:1999xf}. The Higgs spectrum of the two Higgs
doublet extension of the SM contains two charged and three neutral
scalar states. Because in the MPP inspired 2HDM CP--invariance is
preserved one of the neutral Higgs bosons is always CP--odd while
two others are CP--even. The charged and pseudoscalar Higgs states
gain masses \be
m^2_{\chi^{\pm}}=m_A^2-\ds\frac{\lambda_4}{2}v^2\,,\qquad\qquad
m_A^2=\frac{2m_3^2}{\sin 2\beta}\,. \label{2hdm51} \ee The direct
searches for the rare B--meson decays ($B\to X_s\gamma$) place a
lower limit on the charged Higgs scalar mass in the 2HDM of type
II \cite{mch}: \be m_{\chi^{\pm}}> 350\,\mbox{GeV}\,,
\label{2hdm52} \ee which is also valid in our case.

The CP--even states are mixed and form a $2\times 2$ mass matrix.
The diagonalisation of this matrix gives \be
m_{h_1,\,h_2}^2=\frac{1}{2}\left(M^2_{11}+M^2_{22}\mp\sqrt{(M_{22}^2-M_{11}^2)^2+4M^4_{12}}\right)\,.
\label{2hdm55} \ee
$$
\ba{rcl}
M_{11}^2&=&\biggl(\lambda_1\cos^4\beta+\lambda_2\sin^4\beta+\ds\frac{\lambda}{2}\sin^22\beta\biggr)v^2\,,\\[2mm]
M_{12}^2&=&M_{21}^2=\ds\frac{v^2}{2}\biggl(-\lambda_1\cos^2\beta+\lambda_2\sin^2\beta+\lambda\cos 2\beta
\biggr)\sin 2\beta\,,\\[2mm]
M_{22}^2&=&m_A^2+\ds\frac{v^2}{4}\biggl(\lambda_1+\lambda_2-2\lambda\biggr)\sin^22\beta\,,
\ea
$$
where $\lambda=\lambda_3+\lambda_4$. The qualitative pattern of
the Higgs spectrum depends very strongly on the mass of the
pseudoscalar Higgs boson $m_A$. With increasing $m_A$ the masses
of all the Higgs particles grow. At very large values of $m_A$
($m_A^2>>v^2$) the lightest Higgs boson mass $m_{h_1}$ approaches
its theoretical upper limit $\sqrt{M_{11}^2}$.

The upper bound on the mass of the lightest CP--even Higgs boson
only depends on the Higgs self--couplings and $\tan\beta$.
Therefore, using the results of our numerical studies of the RG
flow presented in Table 1, one can calculate the theoretical
restriction on $m_{h_1}$ near the quasi--fixed point for each
value of the MPP scale. The direct computations demonstrate that
the allowed interval of variation of the lightest Higgs boson mass
enlarges when $\Lambda$ approaches the electroweak scale. The
increase in the upper bound on $m_{h_1}$ is caused by the growth
of $\lambda_i(M_t)$ in the vicinity of the quasi--fixed point. In
Fig.~3a we plot the theoretical restriction on the lightest Higgs
boson mass $m_{h_1}$ in the MPP inspired 2HDM as a function of
scale $\Lambda$ for $h_t^2(\Lambda)=10$ and $h_t^2(\Lambda)=2.25$.
Fig.~3a illustrates that at high scale $\Lambda$ the upper bound
on $m_{h_1}$ grows slightly with decreasing $h_t(\Lambda)$. When
$\Lambda\simeq M_{Pl}$ the variation of $h_t^2(\Lambda)$ from $10$
to $2.25$ raises the theoretical limit on the mass of the SM--like
Higgs boson from $110\,\mbox{GeV}$ to $120\,\mbox{GeV}$. This
indicates that in the MPP inspired 2HDM the scenarios with high
scales $\Lambda$ and large values of $h_t^2(\Lambda),
\lambda_1(\Lambda)$ and $\lambda_2(\Lambda)$ have not been
entirely ruled out by unsuccessful Higgs searches at LEP. If
$\Lambda\gtrsim 10^{10}\,\mbox{GeV}$ the lightest CP--even Higgs
boson is lighter than $125\,\mbox{GeV}$. The upper bound on
$m_{h_1}$ grows from $125\,\mbox{GeV}$ to $140\,\mbox{GeV}$ on
lowering the MPP scale from $10^{10}\,\mbox{GeV}$ to
$10^{7}\,\mbox{GeV}$ (see Fig.~3a and Table 1).

Stringent constraints coming from the direct Higgs searches at LEP
suggest that the spectrum of Higgs bosons should be analysed
together with the Higgs couplings to the gauge bosons and quarks.
Such an analysis is especially important in the LHC era, because
the same couplings determine the production cross sections and
branching ratios of the Higgs particles at the LHC. Following the
traditional notations we define normalised $R$--couplings of the
neutral Higgs states to vector bosons as follows:
$g_{VVh_i}=R_{VVh_i}\times$ SM coupling
$\left(\mbox{i.e.}\,\,\dfrac{\bar{g}}{2}M_V\right)$;
$g_{ZAh_i}=\ds\frac{\bar{g}}{2}R_{ZAh_i}$, where $V$ is a
$W^{\pm}$ or a $Z$ boson. The relative couplings $R_{ZZh_i}$ and
$R_{ZAh_i}$ are given in terms of the angles $\alpha$ and $\beta$
\cite{Carena:2002es}: \be \ba{c}
R_{ZZh_1}=R_{WWh_1}=-R_{ZAh_2}=\sin(\beta-\alpha)\,,\\
R_{ZZh_2}=R_{WWh_2}=R_{ZAh_1}=\cos(\beta-\alpha)\,, \ea
\label{2hdm57} \ee where the angle $\alpha$ is defined as follows:
\be \ba{rcl}
h_1&=&-(H_1^0-v_1)\sin\alpha+(H_2^0-v_2)\cos\alpha\,,\\[3mm]
h_2&=&(H_1^0-v_1)\cos\alpha+(H_2^0-v_2)\sin\alpha\,,
\ea
\label{2hdm58}
\ee
$$
\tan\alpha=\dfrac{(\lambda v^2-m_A^2)\sin\beta\cos\beta}{m_A^2\sin^2\beta+\lambda_1 v^2\cos^2\beta-m_{h_1}^2}\,.
$$
The absolute values of the $R$--couplings $R_{VVh_i}$ and
$R_{ZAh_i}$ vary from zero to unity.

The couplings of the Higgs eigenstates to the top quark
$g_{t\bar{t}h_i}$ can also be presented as a product of the
corresponding SM coupling and the $R$--coupling $R_{t\bar{t}h_i}$:
\be
R_{t\bar{t}h_1}=\dfrac{\cos\alpha}{\sin\beta}\,,\qquad\qquad\qquad
R_{t\bar{t}h_2}=\dfrac{\sin\alpha}{\sin\beta}\,. \label{2hdm61}
\ee Since the $R_{t\bar{t}h_i}$ are inversely proportional to
$\sin\beta$ and near the quasi--fixed point $\tan\beta\lesssim 1$
(see Table 1), the values of $R_{t\bar{t}h_i}$ can be
substantially larger than unity.

As follows from Eqs.~(\ref{2hdm51})--(\ref{2hdm61}), the spectrum
and couplings of the Higgs bosons in the MPP inspired 2HDM are
parametrized in terms of $m_A$, $\tan\beta$ and four Higgs
self--couplings $\lambda_1(M_t),\,
\lambda_2(M_t),\,\lambda_3(M_t)$ and $\lambda_4(M_t)$. Near the
quasi--fixed points the Higgs self--couplings, the top quark
Yukawa coupling and $\tan\beta$ have already been calculated (see
Table 1). The numerical values of these couplings depend on the
MPP scale. Therefore in the quasi--fixed point scenario all the
Higgs masses and couplings can be considered as functions of the
scale $\Lambda$ and pseudoscalar mass $m_A$ only.

In Fig.~3b--3d we examine the dependence of the Higgs masses and
couplings on $m_A$ for the MPP scale $\Lambda=10\,\mbox{TeV}$.
From Fig.~3b it becomes clear that the masses of the heaviest
CP--even and charged Higgs states rise with increasing
pseudoscalar mass. At large values of $m_A$ the corresponding
Higgs states are almost degenerate around $m_A$. The lightest
Higgs scalar $h_1$ is predominantly a SM--like Higgs boson,
because its relative coupling to a $Z$ pair is always close to
unity (see Fig.~3c). As a result the non-observation of the
SM--like Higgs particle at LEP rules out most of the parameter
space near the quasi--fixed point if the scale $\Lambda$ is
relatively high, i.e. $\Lambda\gtrsim 10^{10}\,\mbox{GeV}$. When
the pseudoscalar mass is large ($m_A\gg M_t$) the interaction of
the lightest CP--even Higgs state with the Higgs pseudoscalar and
$Z$ is suppressed.

The relative couplings of the CP--even Higgs bosons to the top
quark change considerably when $\Lambda$ varies. When $\Lambda$ is
near the Planck scale the lightest CP--even Higgs eigenstate is
predominantly composed of $H_1^0$. Therefore its coupling to the
top quark is typically smaller than the coupling of the heaviest
one. However at low values of the MPP scale, $\Lambda <
10^6\,\mbox{GeV}$, the lightest CP--even Higgs state is dominated
by $H_2^0$. As follows from Fig.~3d this leads to a substantial
increase of the coupling of the lightest Higgs scalar to the top
quark. Our numerical studies demonstrate that, due to the
significant growth of $R_{t\bar{t}h_1}$, the production cross
section of the SM--like Higgs in the 2HDM can be $1.5-2$ times
larger than in the SM. With increasing $m_A$ the heaviest
CP--even, CP--odd and charged Higgs states decouple and the
coupling of the lightest Higgs scalar to the top quark approaches
the SM one. Nevertheless the enhanced production of the SM--like
Higgs boson allows us to distinguish the quasi--fixed point
scenario in the MPP inspired 2HDM with low MPP scale from the SM
and its supersymmetric extensions, even if extra Higgs states are
relatively heavy ($m_A\gtrsim 500-700\,\mbox{GeV}$).

\section{Conclusions}

We have studied the RG flow of $h_t(\mu)$ and $\lambda_i(\mu)$, as
well as the Higgs spectrum and couplings, within the simplest two
Higgs doublet extension of the SM --- the MPP inspired 2HDM. When
$h_t(\Lambda)\gtrsim 1$ the solutions of the RG equations for the
top quark Yukawa coupling are concentrated in the vicinity of the
quasi--fixed point at the electroweak scale. Then the value of
$\tan\beta$ can be chosen so that the correct value of the running
top quark mass is reproduced. In the MPP inspired 2HDM the values
of $h_t(M_t)$ and $\tan\beta$ corresponding to the quasi--fixed
point depend mainly on the MPP scale $\Lambda$. We have argued
that, at large values of $h_t(\Lambda)$, the MPP and vacuum
stability conditions constrain the Higgs self--couplings at the
MPP scale very strongly. When the scale $\Lambda$ is high enough
the $\lambda_i(\Lambda)$ are confined in a narrow region near a
point corresponding to a fixed point of the RG equations in the
gaugeless limit ($g_i=0$). This ensures the convergence of the
solutions for $\lambda_1(\mu)$ and $\lambda_2(\mu)$ to the
quasi--fixed points. Two other non--zero Higgs self--couplings
$\lambda_3(\mu)$ and $\lambda_4(\mu)$ are attracted considerably
weaker to the quasi--fixed points.

The qualitative pattern of the Higgs spectrum in the MPP inspired
2HDM is determined by the mass of the pseudoscalar Higgs boson
$m_A$. When $m_A\gg M_t$ the masses of the charged, CP--odd and
heaviest CP--even Higgs bosons are almost degenerate around $m_A$.
In the considered limit the lightest CP--even Higgs boson mass
$m_{h_1}$ attains its maximal value. Using the results of our
analysis of the RG flow of the Higgs and Yukawa couplings, we have
examined the dependence of the upper bound on $m_{h_1}$ near the
quasi--fixed point on the scale $\Lambda$. If $\Lambda\gtrsim
10^{10}\,\mbox{GeV}$ the mass of the lightest Higgs particle does
not exceed $125\,\mbox{GeV}$. However at low MPP scale
$\Lambda\simeq 10-100\,\mbox{TeV}$ the upper bound on $m_{h_1}$
reaches $200-220\,\mbox{GeV}$. The lightest Higgs scalar in the
considered case is predominantly a SM--like Higgs boson, since its
relative coupling to a $Z$ pair is rather close to unity.
Nevertheless at low MPP scales the quasi--fixed point scenario
leads to large values of the relative coupling of the lightest
Higgs scalar to the top quark, resulting in the enhanced
production of this particle at the LHC. Thus the analysis of
production and decay rates of the SM--like Higgs boson at the LHC
should make possible the distinction between the quasi--fixed
point scenario in the MPP inspired 2HDM with low scale $\Lambda$,
the SM and the MSSM even if extra Higgs states are relatively
heavy, i.e. $m_A\simeq 500-700\,\mbox{GeV}$.

\section*{Acknowledgements}

\vspace{-2mm} The authors are grateful to L.~V.~Laperashvili and M.~Sher for valuable comments and remarks.
RN would also like to thank A.~Djouadi, J.~F.~Gunion, D.~J.~Miller, J.~Kalinowski, D.~I.~Kazakov, S.~F.~King,
M.~Shifman and P.~M.~Zerwas for fruitful discussions. The authors acknowledge support from the SHEFC grant
HR03020 SUPA 36878.

\newpage

\begin{center}
{\bf Table 1.} The top quark Yukawa and Higgs couplings,
$\tan\beta$ and the upper bound on the lightest Higgs boson mass
corresponding to
the quasi--fixed point scenario in the MPP inspired 2HDM (all mass parameters are given in GeV).\\
\vspace{5mm}
\begin{tabular}{|c|c|c|c|c|c|c|c|}
\hline
$\Lambda$ & $\hat{h}_t(M_t)$ & $\tan\beta$ & $\hat{\lambda}_1(M_t)$  & $\hat{\lambda}_2(M_t)$
& $\hat{\lambda}_3(M_t)$ & $\hat{\lambda}_4(M_t)$ & $m_{h_1}$\\
 \hline
$M_{Pl}$  & $1.26$ & $1.08$ & $0.41$ & $0.94$ & $0.037$ & $-0.33$ & $114$ \\
\hline
$10^{16}$ & $1.30$ & $1.02$ & $0.48$ & $1.02$ & $0.038$ & $-0.36$ & $115$ \\
\hline
$10^{13}$ & $1.36$ & $0.94$ & $0.57$ & $1.15$ & $0.035$ & $-0.40$ & $118$ \\
\hline
$10^{10}$ & $1.45$ & $0.84$ & $0.73$ & $1.36$ & $0.019$ & $-0.49$ & $124$ \\
\hline
$10^{7}$  & $1.61$ & $0.71$ & $1.05$ & $1.78$ & $-0.057$ & $-0.67$ & $143$ \\
\hline
$10^{4}$  & $2.05$ & $0.51$ & $2.09$ & $3.09$ & $-0.65$ & $-1.21$ & $226$ \\
\hline
\end{tabular}
\end{center}

\newpage
\begin{figure}
\hspace*{-0cm}{$h_t(\mu)$}\hspace*{7.7cm}{$h_t(\mu)$}\\[1mm]
\includegraphics[height=52mm,keepaspectratio=true]{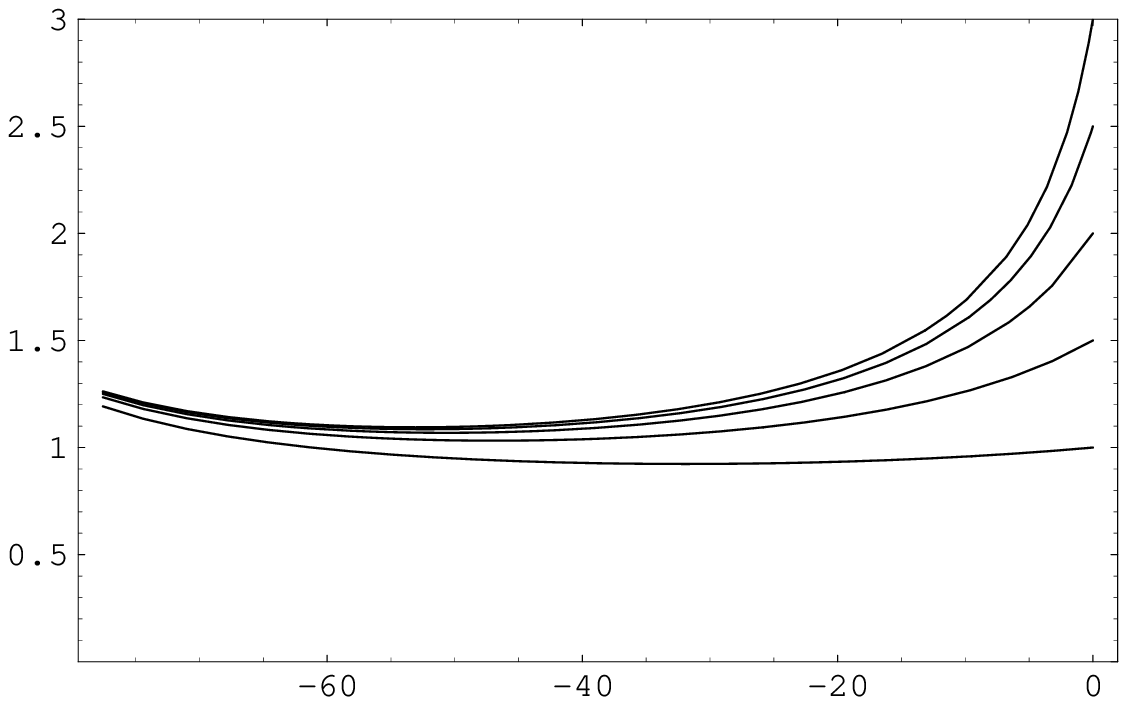}
\includegraphics[height=52mm,keepaspectratio=true]{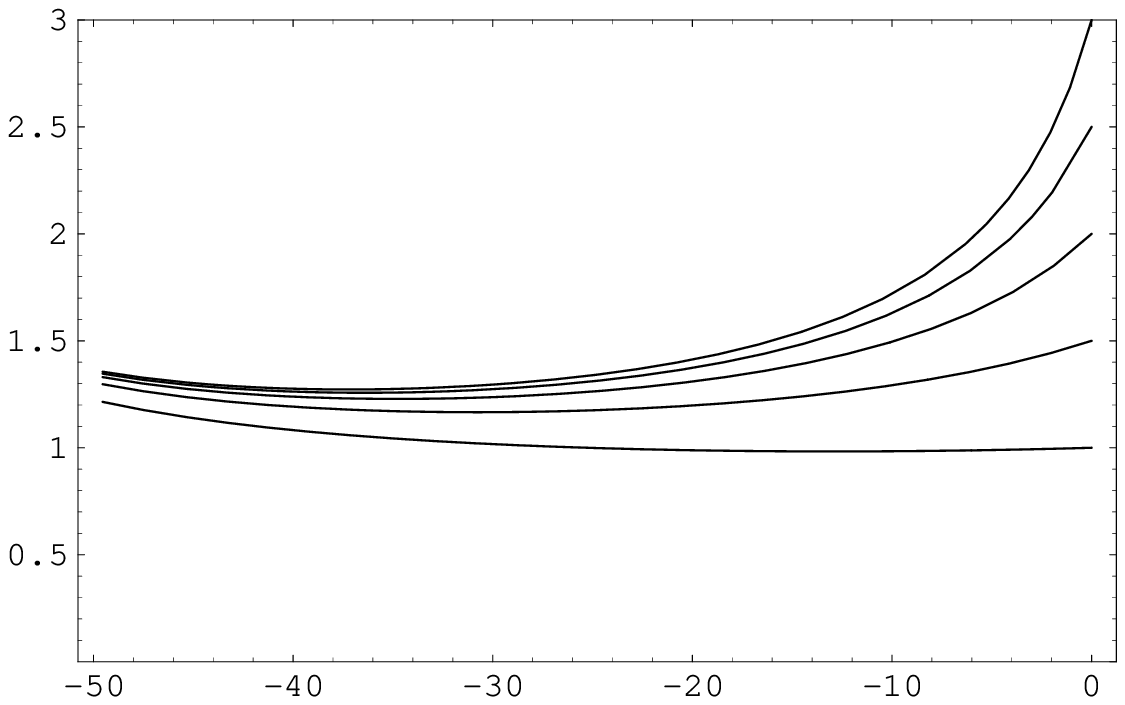}\\
\hspace*{3.5cm}{$\log[\mu^2/M_{Pl}^2]$}\hspace*{6.5cm}{$\log[\mu^2/\Lambda^2]$}\\[1mm]
\hspace*{4.5cm}{\bf (a)}\hspace*{7.5cm}{\bf (b) }\\
\caption{The RG flow of the top quark Yukawa coupling for {\it
(a)} $\Lambda=M_{Pl}$ and {\it (b)} $\Lambda=10^{13}\,\mbox{GeV}$.
The value of $\alpha_3(M_Z)$ is set equal to $0.117$.}
\label{2hdm1}
\end{figure}

\begin{figure}
\hspace*{-0cm}{$\lambda_1(\mu)$}\hspace*{7.7cm}{$\lambda_2(\mu)$}\\[1mm]
\includegraphics[height=52mm,keepaspectratio=true]{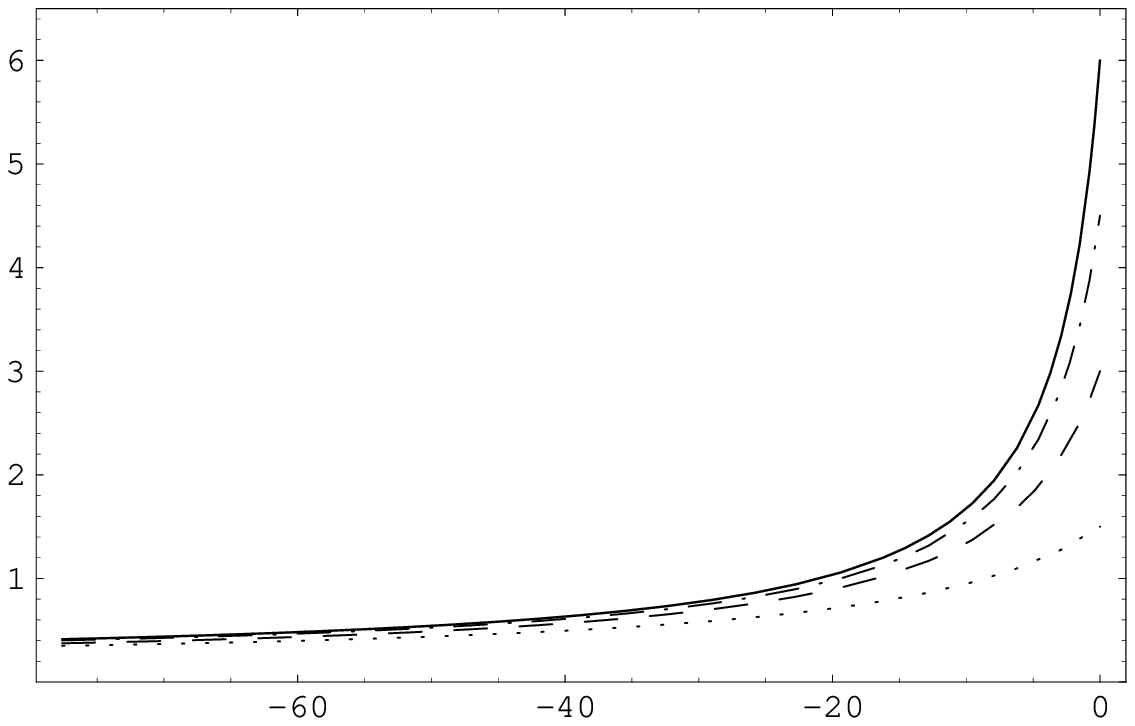}
\includegraphics[height=52mm,keepaspectratio=true]{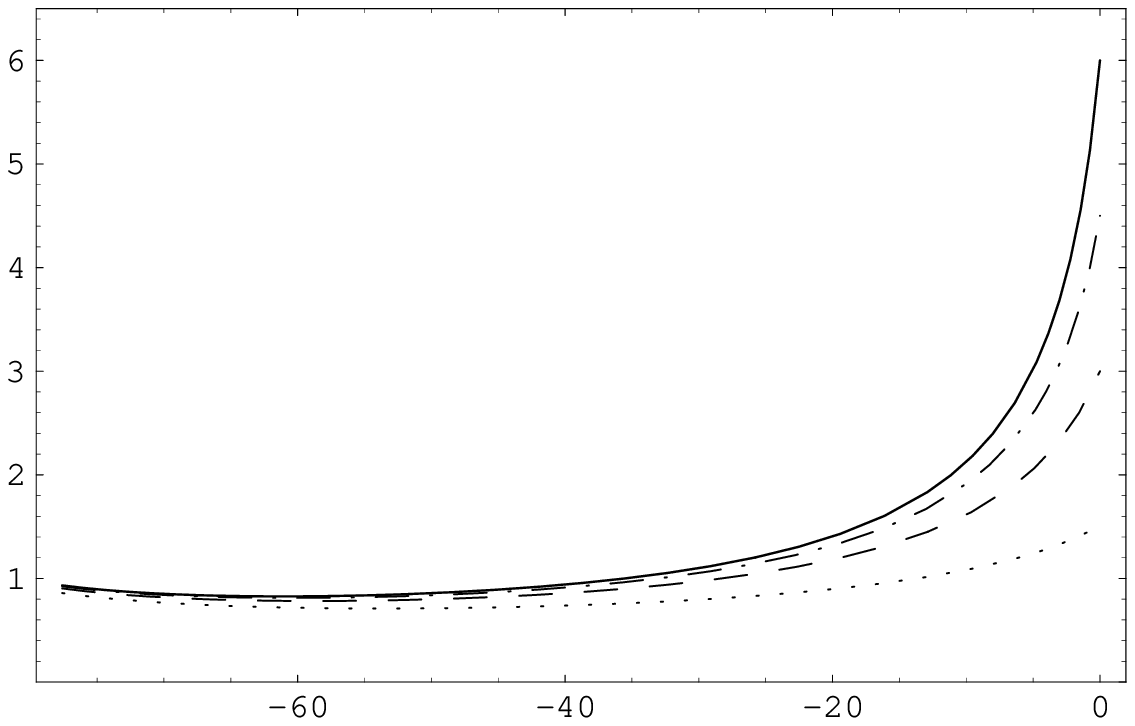}\\
\hspace*{3.5cm}{$\log[\mu^2/M_{Pl}^2]$}\hspace*{7.5cm}{$\log[\mu^2/M_{Pl}^2]$}\\[1mm]
\hspace*{4.5cm}{\bf (a)}\hspace*{7.5cm}{\bf (b) }\\[3mm]
\hspace*{-0cm}{$\lambda_3(\mu)$}\hspace*{7.7cm}{$\lambda_4(\mu)$}\\[1mm]
\includegraphics[height=52mm,keepaspectratio=true]{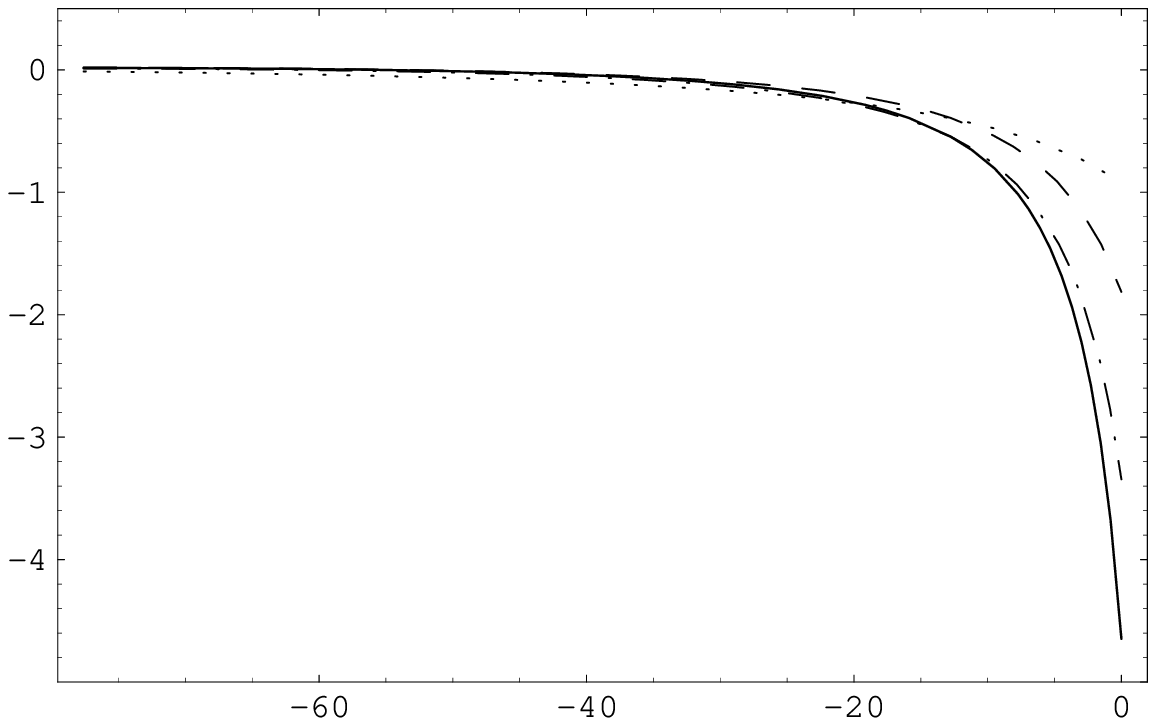}
\includegraphics[height=52mm,keepaspectratio=true]{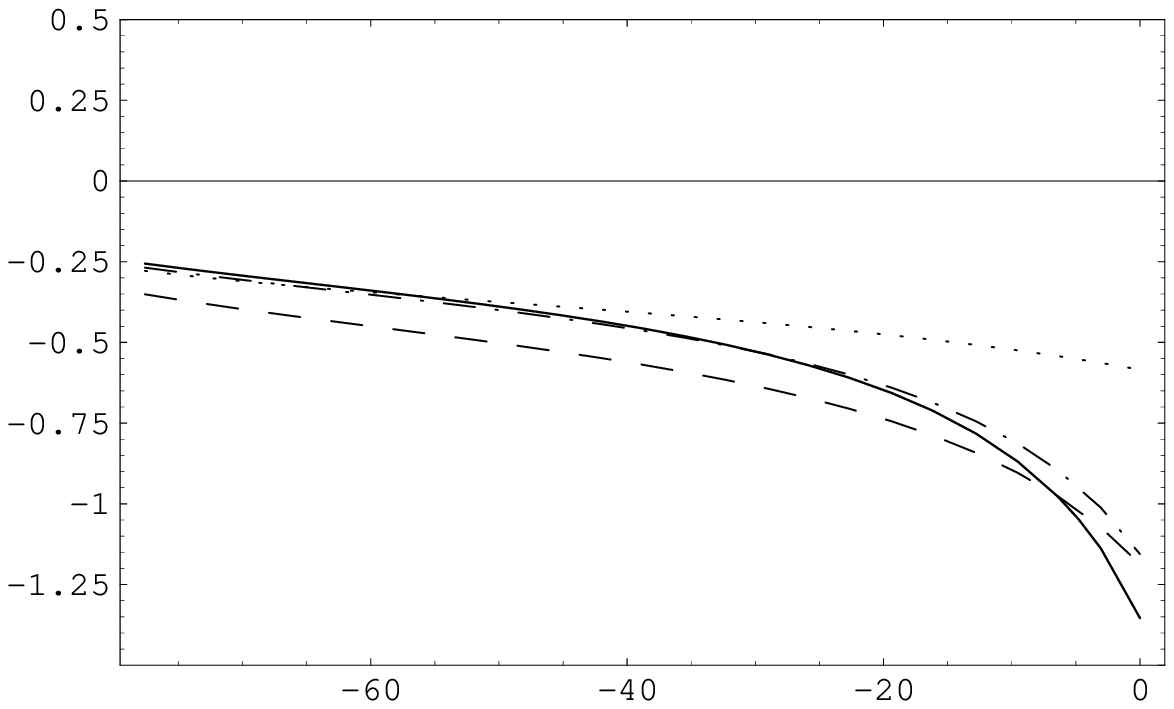}\\
\hspace*{3.5cm}{$\log[\mu^2/M_{Pl}^2]$}\hspace*{7.5cm}{$\log[\mu^2/M_{Pl}^2]$}\\[1mm]
\hspace*{4.5cm}{\bf (c)}\hspace*{7.5cm}{\bf (d) }\\
\caption{The RG flow of the Higgs self--couplings within the MPP
inspired 2HDM: {\it (a)} $\lambda_1(\mu)$; {\it (b)}
$\lambda_2(\mu)$; {\it (c)} $\lambda_3(\mu)$; {\it (d)}
$\lambda_4(\mu)$. The solid, dash--dotted, dashed and dotted lines
correspond to different sets of $(h_t(\Lambda),\,
\lambda_1(\Lambda)=\lambda_2(\Lambda)=\lambda_0)$, i.e.
$(2.65,\,6)$, $(2.3,\,4.5)$, $(1.9,\,3)$ and $(1.35,\,1.5)$
respectively, while $\lambda_3(\Lambda)$ and $\lambda_4(\Lambda)$
are chosen so that the MPP conditions are fulfilled.}
\label{2hdm2}
\end{figure}

\begin{figure}
\hspace*{-0cm}{$m_{h_1}$}\hspace*{7.7cm}{$m_{h_i,\,\chi^{\pm}}$}\\[1mm]
\includegraphics[height=52mm,keepaspectratio=true]{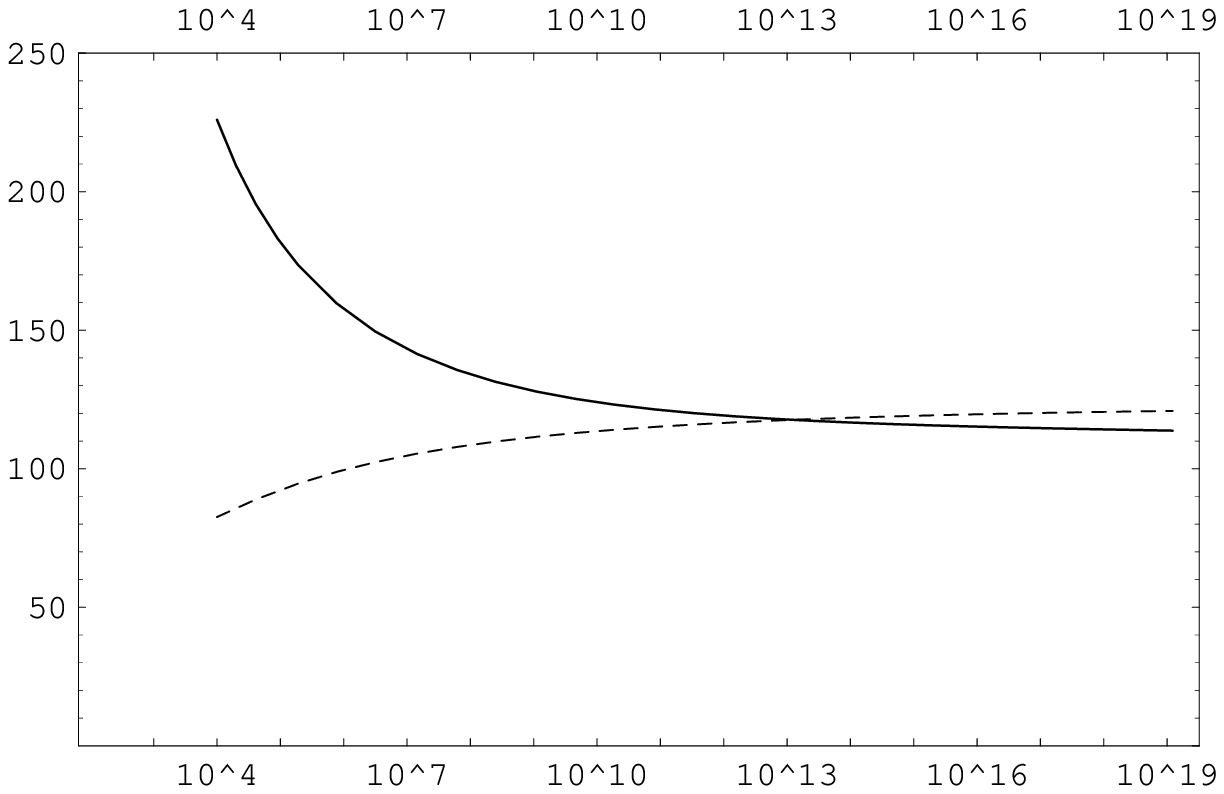}
\includegraphics[height=52mm,keepaspectratio=true]{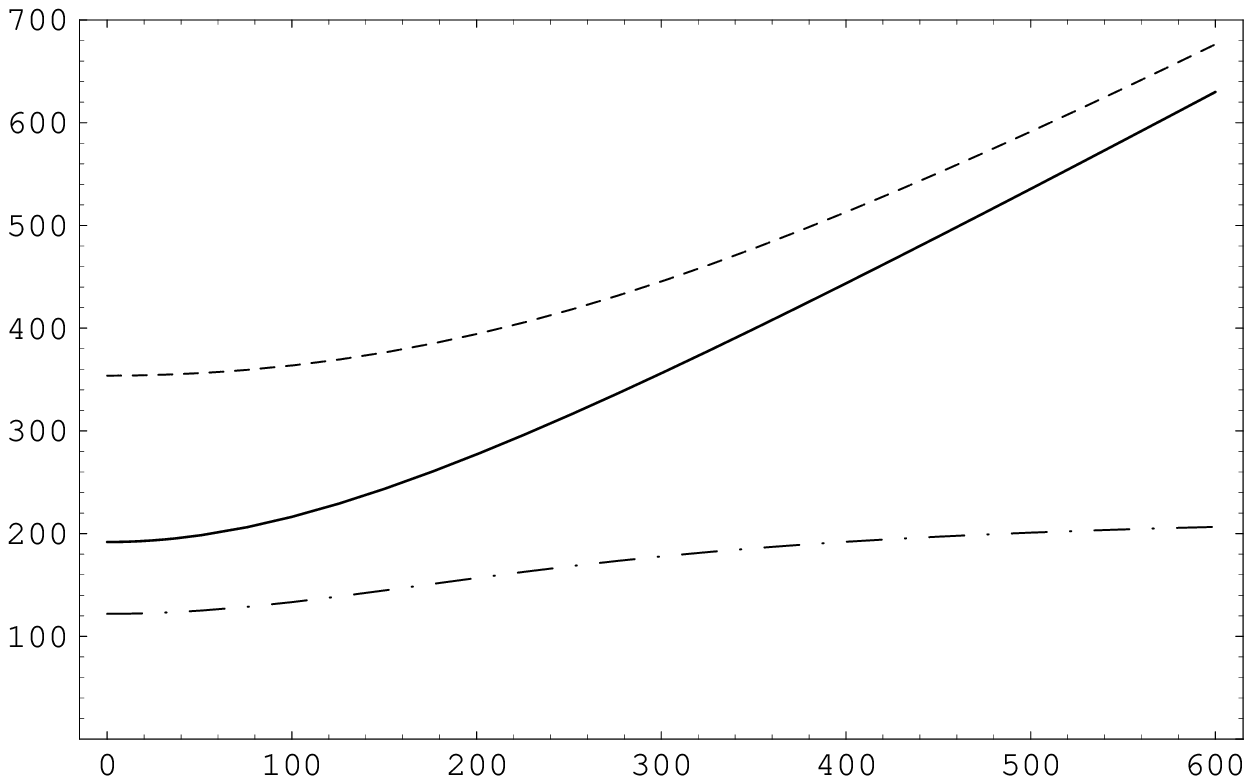}\\
\hspace*{4.5cm}{$\Lambda$}\hspace*{7.5cm}{$m_A$}\\[1mm]
\hspace*{4.5cm}{\bf (a)}\hspace*{7.5cm}{\bf (b) }\\[3mm]
\hspace*{-0cm}{$|R_{ZZh_i}|$}\hspace*{7.7cm}{$|R_{t\bar{t}h_i}|$}\\[1mm]
\includegraphics[height=52mm,keepaspectratio=true]{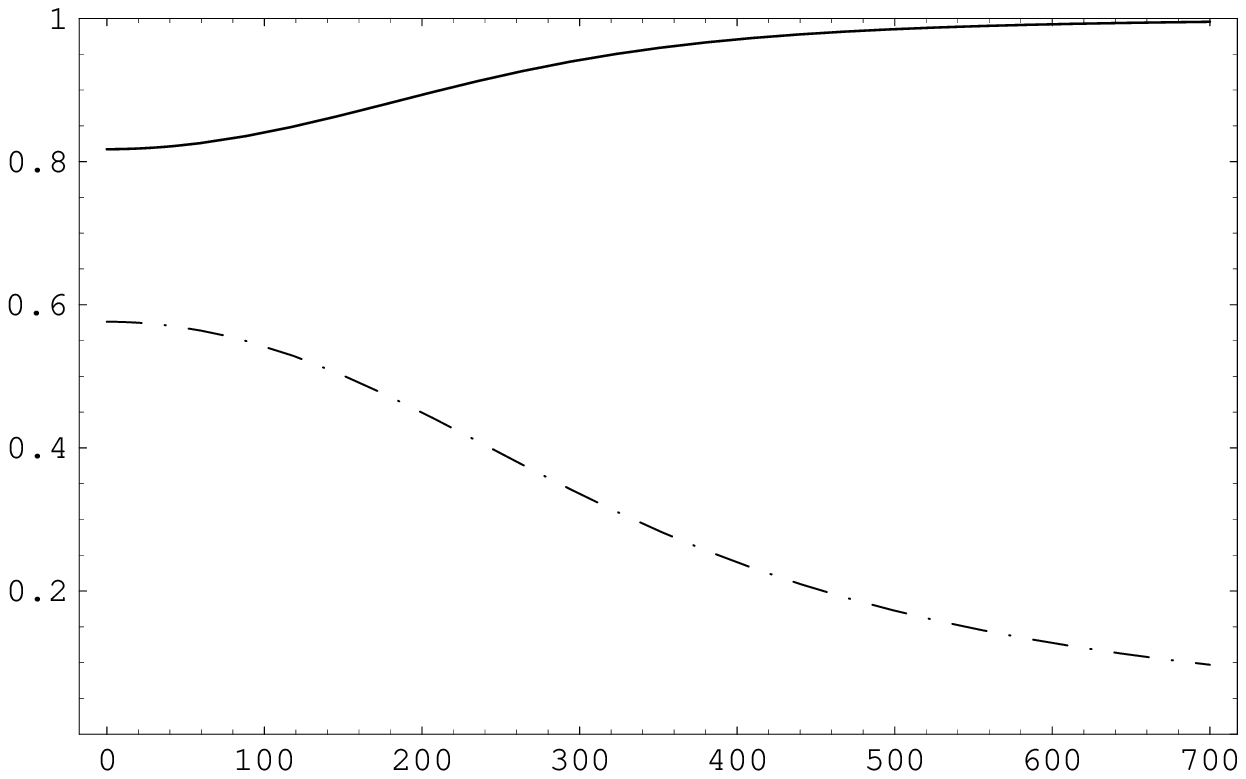}
\includegraphics[height=52mm,keepaspectratio=true]{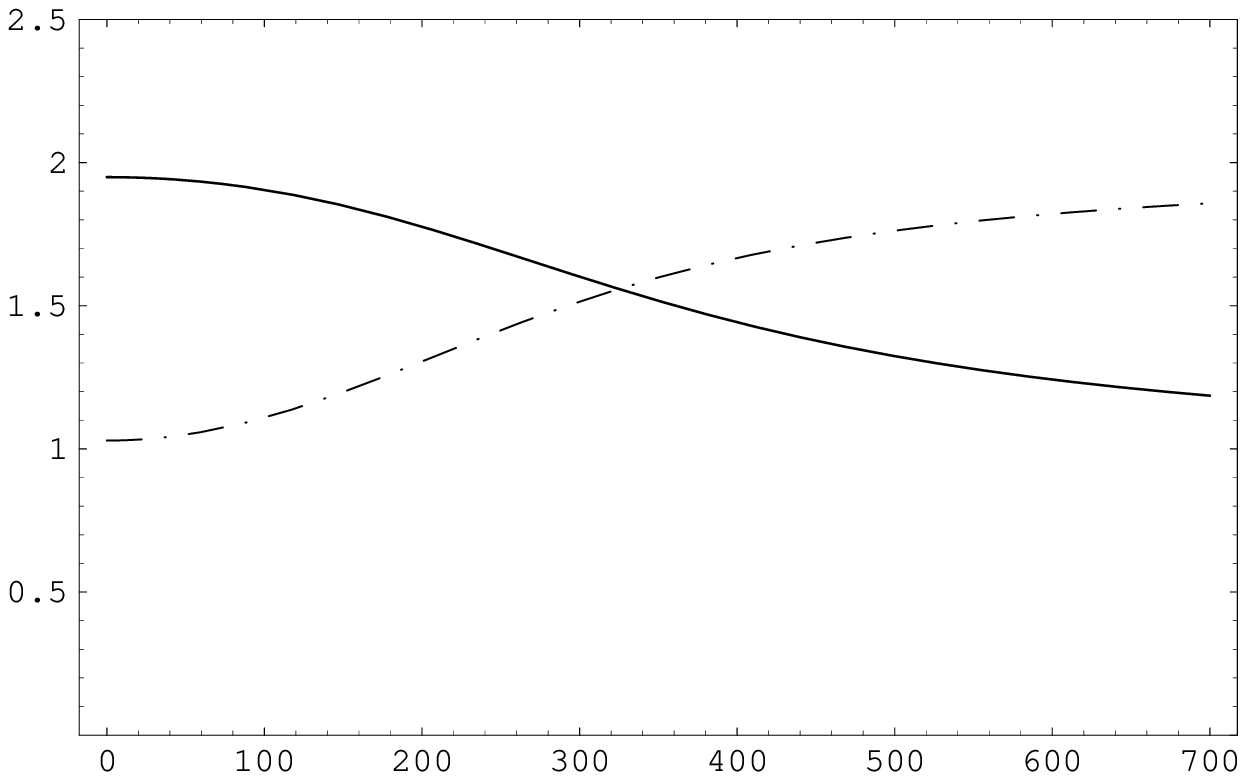}\\
\hspace*{4.5cm}{$m_A$}\hspace*{7.5cm}{$m_A$}\\[1mm]
\hspace*{4.5cm}{\bf (c)}\hspace*{7.5cm}{\bf (d) }\\
\caption{Higgs masses and couplings near the quasi--fixed point in
the MPP inspired 2HDM. {\it (a)} Upper bound on the mass of the
SM--like Higgs boson versus MPP scale $\Lambda$ in the
quasi--fixed point scenario. The solid and dashed curves
correspond to $h_t^2(\Lambda)=10$ and $h_t^2(\Lambda)=2.25$. {\it
(b)} Spectrum of Higgs bosons versus $m_A$ for
$\Lambda=10\,\mbox{TeV}$ and $h^2_t(\Lambda)=10$. The dash--dotted
and dashed lines correspond to the CP--even Higgs boson masses,
while the solid line represents the mass of the charged Higgs
states. {\it (c)} Absolute values of the relative couplings
$R_{ZZi}$ of the Higgs scalars to a $Z$ pair. Solid and
dashed--dotted curves represent the dependence of the couplings of
the lightest and heaviest CP--even Higgs states to a $Z$ pair on
$m_A$. The parameters are the same as in {\it (b)}. {\it (d)}
Absolute values of the relative couplings $R_{t\bar{t}i}$ of the
lightest (solid curve) and heaviest (dashed--dotted curve)
CP--even Higgs bosons to the top quark as a function of $m_A$. The
parameters are the same as in {\it (b)}--{\it (c)}. All mass
parameters are given in GeV.} \label{2hdm3}
\end{figure}

\end{document}